\documentclass[universe,article,accept,moreauthors,pdftex]{Definitions/mdpi} 
\firstpage{1} 
\pubvolume{1}
\issuenum{1}
\articlenumber{0}
\pubyear{2026}
\copyrightyear{2026}
\datereceived{} 
\dateaccepted{} 
\datepublished{} 
\hreflink{https://doi.org/} 

\Title{Superfluid Vortex Lines-Magnetic Flux Tubes Interaction and Braking Indices of Pulsars}

\TitleCitation{...}

\Author{Erbil G\"{u}gercino\u{g}lu$^{1*}$}

\AuthorNames{Erbil G\"{u}gercino\u{g}lu}

\AuthorCitation{G\"{u}gercino\u{g}lu, E.}

\address{%
$^{1}$ \quad School of Arts and Sciences, Qingdao Binhai University, 266555 Qingdao, People's Republic of China; egugercinoglu@gmail.com}

\corres{Correspondence: egugercinoglu@gmail.com;  (E.G.)}

\abstract{
Braking of pulsars, or the law of their spin deceleration, is a manifestation of the combination of various processes occurring in the magnetospheres of neutron stars and their internal structural dynamics. The interaction of superfluid neutron vortex lines with superconducting magnetic flux tubes in the neutron star core plays a significant role in rotational evolution and magnetic field evolution through spin glitches and magnetic flux expulsion. In this study, the effect of this interaction on the temporal variation of the braking index of pulsars is investigated. The variation of the deviation from the $n=3$ value predicted by the generally accepted magnetic dipole formula with respect to physical parameters such as the magnetic field and the age of the neutron star has been elaborated, and applications have been made to pulsars with reliably measured braking indices.
}

\keyword{neutron stars; magnetic field; radio pulsars; superconductivity; superfluidity} 

\begin{document}

\section{Introduction} 
The long-term rotational evolution of the pulse frequency $\nu$ of a rotation-powered pulsar, which is expressed as a power law of the form $\dot{\nu}=-K\nu^{n}$ along with a structural constant $K$ corresponding to an unspecified braking torque, provides invaluable insights into the processes governing its magnetosphere and internal dynamics, as well as a way to extract various types of noise from timing data \cite{vargas23,vargas24,vargas25}. In young pulsars, high braking indices in the interval $n=10-100$ have been measured, reflecting the domination of the inter-glitch periods of pulsars by superfluid internal torques rather than external magnetospheric torques \cite{alpar06,lower21,erbil26}. In older pulsars, both positive and negative values of $\ddot{\nu}$ are observed from approximately the same number of sources for each sign. \citep{johnston99,hobbs10,ou16,parthasarathy20}. This observation can be explained by either a fluctuating component in the long-term magnetic field evolution \cite{biryukov12,zhang12} or a time variable internal torque on the neutron star crust due to superfluid vortex bending oscillations \cite{erbil23}. 

For mature pulsars whose spin evolution is not interrupted by rotational glitches and whose second time derivative of frequency $\ddot{\nu}$ changes so greatly that it can be measured in periods on the order of a decade or shorter, the braking index $n$ can be reliably determined if the magnetospheric noise component can also be sorted out. For such pulsars, the range $1 \lesssim n \lesssim 3$ has been observed, and very different physical mechanisms could be responsible for this; for a review, see Ref.\cite{abolmasov24}. The present study attempts to understand the braking indices of radio pulsars within the framework of the interaction between superfluid vortex lines and magnetic flux tubes, based on a model concerning the superfluid-superconducting internal structure of neutron stars.

This paper is organized as follows. Section \ref{sfn-scp} provides an outline of the superfluid and superconducting dynamics of neutron stars, in relation to the calculation of vortex line and flux tube velocities in the liquid core. Section \ref{vl-ft} lays the foundation for the vortex line-flux tube interaction and discusses the correct form of the magnetic force density from which the flux tube velocity is evaluated. Section \ref{app-n} introduces the vortex line-flux tube interaction model and presents an application to the observed data on the braking indices of radio pulsars. And finally, Section \ref{dis-conc} summarizes the findings and results with a conclusion on future prospects.

\section{Overview of Neutron Star Superfluidity and Superconductivity}
\label{sfn-scp}

A few hundred years after its formation, a rapidly rotating and highly magnetized neutron star is expected to have a liquid core of $^{1}S_{0}$ superconducting protons, $^{3}P_{2}$ superfluid neutrons, and extremely relativistic and degenerate electrons, which are surrounded by a thin highly conducting solid crust consisting of neutron-rich nuclei and relativistic degenerate electrons embedded in a $^{1}S_{0}$ neutron superfluid liquid \cite{sauls89,pines85,chamel17,haskell18}. These quantum fluids are threaded by a dense array of quantized vortex lines, which can interact strongly with coexisting lattice nuclei in the inner crust \cite{alpar84,seveso16} and  quantized magnetic flux-tubes in the outer core \cite{ruderman98,erbil14}. Such interactions play an important role in the magneto-thermal-rotational evolution of pulsars and drive various phenomena \cite{lamb91,geppert17}.

After transitioning to the superfluid phase, the superfluid neutrons of a neutron star can achieve rigid rotation by establishing an array of quantized vortex lines parallel to the stellar spin axis, with an areal density
\begin{equation}
    n_{\rm v}=\frac{2\Omega}{\kappa}\cong6\times10^{3}P(\rm{s})^{-1} \,\mbox{cm$^{-2}$},
    \label{vorden}
\end{equation}
where $\Omega=2\pi/P$ is the angular speed of the pulsar ($P=1/\nu$ is the stellar spin period) and $\kappa=h/2m_{\rm n}\cong2\times10^{-3}$ cm\,s$^{-1}$ is the quantized vorticity attached to individual vortex lines with $h$ and $m_{\rm n}$ being Planck's constant and the bare neutron mass, respectively. In the neutron star crust, vortex lines can scatter off the electron cloud surrounding nuclei \cite{bildsten89}, lattice defects \cite{jones98} and crystal phonons \cite{epstein92}. The pinning of vortex lines to crustal nuclei is considered as the main mechanism for the accumulation of excess angular momentum that triggers glitches in the pulsar rotation rate. \cite{anderson75,alpar84}. Vortex lines can also pin to  magnetic flux tubes due to neutron-proton velocity difference and density fluctuations at the center of a vortex core \cite{muslimov85,srinivasan90} or more effectively due to magnetic flux endowed with a vortex by the entrained superconducting proton current around the lines \cite{mendell91,glampedakis11}. In order to keep up with the rotation rate of the crust, the vortex lines in a spinning down (up) neutron star must move radially outward (inward) with a velocity
\begin{equation}
  v_{\rm v}=r_{\perp}\frac{\dot{P}}{2P}=\frac{r_{\perp}}{4\tau_{\rm sd}},
  \label{vorvel}
\end{equation}
where $r_{\perp}$ is the distance of the vortex line from the spin axis and $\tau_{\rm sd}=P/2\dot{P}$ is the spin-down time scale of the pulsar. For the flux tubes at the crust base, which are responsible for surface magnetic field changes, the radial distance of the vortices to the rotation axis is of the order of the radius of the neutron star, $r_{\perp}\sim R_{*}$.

Any magnetic field that passes through the neutron star's superconducting core is expected to confine within very inhomogeneously structured, twisted, and bundled flux tubes due to the considerably high electrical conductivity of the liquid core \cite{baym69}. The number of flux tubes per unit area is given by 
\begin{equation}
    n_{\Phi}=\frac{B}{\Phi_{0}}\cong5\times10^{18}\left(\frac{B}{10^{12}\mbox{G}}\right)\,\mbox{cm$^{-2}$},
    \label{fluxden}
\end{equation}
where $B$ is the stellar magnetic field and $\Phi_{0}=hc/2e=2.07\times10^{-7}$\,G cm$^{2}$ is the quantum of magnetic flux with $e$ being the electronic charge unit and $c$ speed of light. Various forces act on flux tubes \cite{ding93,jahan-miri00}. Among these, the pinning interaction with vortex lines and the drag against the electron-proton plasma are the most prominent factors that influence the motion of the flux tubes \cite{ruderman98,jones06,bransgrove18,gusakov19,sourie20}. 

Any magnetic field force density $f_{\rm Mag}$ per unit volume acting on flux tubes would force them to move through the core's electron-proton plasma with a characteristic velocity \cite{ruderman98}
\begin{equation}
    v_{\Phi}\sim\frac{f_{\rm Mag}c^{2}}{\sigma n_{\Phi}^{2}\Phi_{0}^{2}}.
    \label{velrud}
\end{equation}
Note that $n_{\Phi}\Phi_{0}=\overline{B}$, i.e. equal to the locally averaged magnetic field. Here, the electrical resistivity $\sigma^{-1}$ is the sum of the two contributions: 
\begin{equation}
    \sigma^{-1}=\left(\frac{e^{2}n_{e}^{2}}{\eta}+\frac{c^{2}\eta}{\Phi_{0}^{2}}\right)^{-1}n_{\Phi},
    \label{elres}
\end{equation}
where $\eta$ is the drag coefficient corresponding to a force per unit length of the flux tube of the form $=\eta v_{\Phi}$ and is given by \cite{harvey86}
\begin{equation}
    \eta=\frac{3\pi\Phi_{0}^{2}e^{2}n_{e}}{64\Lambda_{*}c E_{\rm F_{\rm e}}},
\end{equation}
where $\Lambda_{*}\approx100$ fm is the London penetration length for the decay of the magnetic flux around a flux tube, $n_{\rm e}$ is the number density of electrons, and $E_{\rm F_{\rm e}}\approx100$ MeV is the electron Fermi energy. The reason for an increase in the electrical resistivity for a superconducting core is that electrons not only collide with protons but also scatter from flux tubes, the latter resulting in much more dissipation. The form of $f_{\rm Mag}$ in equation (\ref{velrud}) that a moving vortex array can strongly push on flux tube entanglement is discussed in the next section.

\section{Vortex Line-Flux Tube Interaction}
\label{vl-ft}

A spinning-down (spinning-up) neutron star's vortex lines must expand (contract) to ensure rotational equilibrium with its surface, which is maintained under the action of magnetospheric decelerating (disk-assisted accelerating) external torques \cite{ghosh07}. Because the flux tubes are far more numerous than the vortex lines, as can be easily seen from equations (\ref{vorden}) and (\ref{fluxden}), when they coincide, the motion of vortex lines effectively pushes the flux tubes towards the crust-core interface unless the flux tube entanglement cannot respond fast enough to take part in the collective vortex motion. The interaction force density $f_{\rm Mag,pin}$ due to pinning of vortex lines to flux tubes is \cite{erbil16,bransgrove18}
\begin{equation}
    f_{\rm Mag,pin}\approx n_{\rm v}\frac{\Phi_{0}^{2}}{8\pi^{2}\Lambda_{*}^{3}}\cong2.4\times10^{21}P(\mbox{s})^{-1}\mbox{dyne cm$^{-3}$}.
    \label{vlftpin}
\end{equation}
Note that these authors corrected a typo in the original vortex line-flux tube pinning force expression used by Ruderman et al.\cite{ruderman98}, which erroneously leads to a factor of $\pi^{2}$ times larger interaction force density. Another aspect of the vortex line-flux tube pinning interaction is that because vortex lines are about 100 times stiffer than flux tubes, at each junction some of the pinning energy is carried over to lengthening of flux tubes \cite{erbil16}. Consequently, the pinning energy is lowered by a factor of 8, and the total reduction with an amount of two orders of magnitude in the magnetic force density considered by Ref.\cite{ruderman98} actually leads to an expulsion velocity of $v_{\rm c}\approx10^{-8}$ cm s$^{-1}$ from equations (\ref{velrud}), (\ref{elres}) and (\ref{vlftpin}) instead of $v_{\rm c}\approx10^{-6}$ cm s$^{-1}$ used in that work, invalidating their numerical results obtained for the braking indices.  

Following Jones \cite{jones06}, we will use the flux tube drift velocity that derives from the Boltzmann collision integral for a total magnetic flux-dependent force density which can be generalized to include the effect of interaction with vortices. The drift velocity of the flux tubes can be approximated as \cite{jones06,bransgrove18}
\begin{equation}
    v_{\Phi}\simeq\frac{\tilde{\sigma}}{4\pi n_{e}^{2}e^{2}}\frac{B H_{\rm c1}}{R_{\rm cur}}\cong2,8\times10^{-8}\left(\frac{3.5\times10^{37} \mbox{cm$^{-3}$}}{n_{e}}\right)^{2}\left(\frac{\tilde{\sigma}}{10^{29} \mbox{s$^{-1}$}}\right)\,\mbox{cm s$^{-1}$},
    \label{fluxvel}
\end{equation}
where $H_{\rm c1}\approx10^{14}$ G is the first critical magnetic field for the appearance of type II superconductivity and $R_{\rm cur}\sim R_{*}$ is the radius of curvature for the flux tubes in the core region. Following Jones \cite{jones06}, we will approximate the electrical conductivity in our calculations as $\tilde{\sigma}\approx10^{29}(B/10^{12} \mbox{G})^{-1}$ s$^{-1}$.
Equation (\ref{fluxvel}) is the flux tube velocity that we shall use in our calculations and applications to the braking index data of radio pulsars below.

\section{Model Application to Braking Index Measurements of Radio Pulsars}
\label{app-n}
The magnitude and age-dependent evolution of the observed braking indices of young radio pulsars can be understood as a consequence of the coupled evolution of the core magnetic field and spin in response to vortex line array-flux tube entanglement interactions.

For a pulsar that slows down only due to magnetic dipole radiation, the spin deceleration law is \cite{spitkovsky06,philippov14}
\begin{equation}
    \dot{\Omega}=-\frac{\mu^{2}\Omega^{3}}{Ic^{3}}\left(1+\sin^{2}{\alpha}\right),
    \label{sdown}
\end{equation}
where, $\mu$ is the magnetic dipole moment and $\alpha$ is the magnetic obliquity angle between the rotation and the magnetic dipole axes of the neutron star, and $I$ is the stellar moment of inertia. Equation (\ref{sdown}) predicts a braking index $n$ between 3 and 3.25 \cite{arzamasskiy15,eksi16}. If, for simplicity, we neglect the time-variation of the obliquity angle, then the braking index of a pulsar can be recast
\begin{equation}
    n=3-\frac{\ddot{P}P}{P^{2}}=\frac{\ddot{\Omega}\Omega}{\dot{\Omega}^{2}}=3-\tau_{\rm sd}\left(\frac{4\dot{\mu}}{\mu}-\frac{2\dot{I}}{I}\right).
    \label{nbrake}
\end{equation}
Following Ruderman et al. \cite{ruderman98}, we make the phenomenological ansatz for the time evolution of the magnetic dipole moment
\begin{equation}
    \frac{\dot{\mu}}{\mu}=k\frac{\dot{\Omega}}{\Omega}\sim\frac{v_{\Phi}}{v_{\rm v}}\left(2\tau_{\rm sd}\right)^{-1},
    \label{mudot}
\end{equation}
where the index $k$ encodes the slope of the trajectory of a pulsar in the period-period derivative ($P-\dot{P}$) diagram, and gives the age dependence for the spin induced magnetic field alternation of pulsars. In equation (\ref{nbrake}) a plausible choice for the time variation of the effective stellar moment of inertia by means of discrete spin-up glitches is $\dot{I}/I=(\Delta\dot{\Omega}_{\rm per}/\dot{\Omega})/t_{\rm ig}$ \cite{erbil20}, with $\Delta\dot{\Omega}_{\rm per}$ being a persistent increase in the deceleration rate remaining from a timing glitch event that does not recover, and $t_{\rm ig}$ is the inter-glitch time until the occurrence of the next spin-up event. For pulsars whose rotational evolutions are not interrupted by abrupt glitch-induced changes, equations (\ref{nbrake}) and (\ref{mudot}) yield 
\begin{equation}
    n\simeq3-2v_{\Phi}/v_{\rm v}.
    \label{napprox}
\end{equation}
Table \ref{table1} summarizes the results of the application of the vortex line-flux tube interaction model to 12 radio pulsars for which braking indices were reliably determined, taking into account timing noise and other timing irregularities. For frequently glitching pulsars shown with an asterisk next to their names, the general expression equation (\ref{nbrake}) was used. For all the remaining cases, the simplified version of equation (\ref{napprox}) was applied. Equation (\ref{vorvel}) was used for the vortex line velocity $v_{\rm v}$, while equation (\ref{fluxvel}) was used for the flux tube velocity $v_{\Phi}$. Regarding $t_{\rm ig}$ the observed average time between two successive large glitches of 1081 d. 2678 d., 2585 d. and 795 d. are used for PSRs B0833$-$45, B1800$-$21, B1823$-$13 and J2229+6114, respectively. From the required $\dot{I}/I$ for glitching pulsars it seems that there is an anti-correlation between $(\Delta\dot{\Omega}_{\rm per}/\dot{\Omega})$ and $\tau_{\rm sd}$. This is consistent with the view that neutron stars develop more high vortex density traps with crustquakes as they age \cite{alpar06,erbil20}.    
\begin{specialtable}[H] 
\tablesize{\scriptsize}
\caption{Braking index measurements of radio pulsars ($n_{\rm obs}$) and vortex line-flux tube interaction model prediction ($n_{\rm model}$). For the sources displaying large and frequent glitches indicated with an asterisk `*' the full expression equation (\ref{nbrake}) is used. For the remaining pulsars the simplified expression equation (\ref{napprox}) is employed. \label{table1}}
\begin{tabular}{lclc}
\toprule
\textbf{Pulsar Name}	& \textbf{$n_{\rm model}$}	& \textbf{$n_{\rm obs}$} & \textbf{Reference}\\
\midrule
B0531+21(Crab)		& 2.54			& 2.51(1)    & \cite{lyne15}\\
B0540$-$69		    & 2.53			& 2.13(1)    & \cite{ferdman15}\\
B0833$-$45(Vela)*	& 2.69			& 2.94(55)   & \cite{grover25}\\
J1119$-$6127		& 2.95			& 2.72(3)    & \cite{antonopoulou15}\\
J1208$-$6238		& 2.92			& 2.598(1)   & \cite{clark16}\\
B1509$-$58		    & 2.86			& 2.832(3)   & \cite{livingstone11}\\
J1734$-$3333		& 2.78			& 0.9(2)     & \cite{espinoza11}\\
B1800$-$21*		    & 2.64			& 1.9(5)     & \cite{espinoza17}\\
B1823$-$13*		    & 2.86			& 2.2(6)     & \cite{espinoza17}\\
J1833$-$1034		& 1.12			& 1.857(1)   & \cite{roy12}\\
J1846$-$0258		& 2.98			& 2.65(1)    & \cite{archibald15}\\
J2229+6114*		    & 2.88		    & 2.63(30)   & \cite{erbil22}\\
\bottomrule
\end{tabular}
\end{specialtable}

\section{Discussion and Conclusions}
\label{dis-conc}
In the core region of a rotating neutron star, superfluid neutron vortex lines interact strongly with an overwhelmingly larger number of magnetic flux tubes. One consequence of such interaction is that the outward motion of core neutron vortices in accordance with spin-down of a neutron star may carry substantial flux tubes with them, which in turn expel magnetic flux out of the core liquid and alter the braking law of pulsars.

In the present study, the vortex line-flux tube interaction model of Ruderman et al. \cite{ruderman98} has been revised by correcting the approximately two-order-of-magnitude overestimate in the pinning force, and updated it by considering the corresponding expression of Jones \cite{jones06}, which also takes into account the screening condition of the magnetic flux for the velocity at which flux tubes move out of the core region. Moreover, we consider the effect of the non-relaxed, persistent increase in rate of deceleration in glitching pulsars, which was neglected in Ruderman et al. \cite{ruderman98}, on the temporal variation of the effective moment of inertia of radio pulsars. Furthermore, the vortex line-flux tube model is applied to a larger sample containing 12 pulsars. 

This simple model, based on a small number of assumptions and having a relatively small number of free parameters, can explain the braking indices of radio pulsars that have been reliably measured over time. As shown in Table \ref{table1}, the agreement between the model predictions and the observed cases is quite satisfactory in most cases. The only observation that seems incompatible with the model is PSR J1734$-$3333. A very low braking index of $n=0.9$ can only be accounted for by the growing of the dipole magnetic field on the surface, which is either diamagnetically shielded or buried by accretion of fall-back disk material after the supernova explosion \citep{güneydas13}.

This study is intended to pave the way for the development of more realistic neutron star models that incorporate other magnetic field decay mechanisms \cite{Igoshev25}, such as ambipolar diffusion, and a self-consistent cooling evolution \cite{rodriguez25}.
Another important implication of the vortex line-flux tube interaction is the build up of magnetic stresses at the crust-core interface along with the changing magnetic field configuration, and the cracking of the neutron star crust, which can lead to pulsar glitches \cite{ruderman98,sedrakian99,zhou22}. The simulation of this intriguing problem using a self-consistent coupled spin-magnetic field evolution model will be left for future work.





\vspace{6pt} 




\acknowledgments{I dedicate this article to the cherished memory of my dear mother, G\"{u}l\"{u}\c{s}an G\"{u}gercino\u{g}lu, who played the biggest role in making me the person I am today and who was my greatest supporter throughout all my academic endeavors.}

\funding{EG is supported by the Doctor Foundation of Qingdao Binhai University (No. BJZA2025025).}

\dataavailability{This theoretical study does not generate any new data.}

\conflictsofinterest{The author declares no conflict of interest.}

\reftitle{References}


\externalbibliography{yes}
\bibliography{bibl}
\end{paracol}
\end{document}